\documentclass[twocolumn,longbibliography,aps,pra,superscriptaddress,showpacs,floatfix]{revtex4-1}
\usepackage{graphics,epsfig,graphicx}
\usepackage{amsbsy,gensymb}
\usepackage{amsmath}
\usepackage{bm}
\usepackage{amsfonts}
\usepackage{cancel}
\usepackage{multirow}
\begin{document}

\title{$N$-boson spectrum from a Discrete Scale Invariance}

\author{A. Kievsky}
\affiliation{Istituto Nazionale di Fisica Nucleare, Largo Pontecorvo 3, 56100 Pisa, Italy}
\author{N. K. Timofeyuk}
\affiliation{Department of Physics, 
University of Surrey, Guildford, Surrey GU2 7XH, United Kingdom}
\author{M. Gattobigio} 
\affiliation{Universit\'e de Nice-Sophia Antipolis, Institut Non-Lin\'eaire de
Nice,  CNRS, 1361 route des Lucioles, 06560 Valbonne, France }

\begin{abstract}
We present the analysis of the $N$-boson spectrum computed using
a soft two-body potential the strength of which has been  varied
in order to cover an extended range of positive and negative values
of the two-body scattering length $a$ close to the unitary limit. The spectrum shows a tree structure 
of two states, one shallow and one deep, attached to the ground-state
of the system with one less particle. It is governed by an unique
universal function, $\Delta(\xi)$, already known in the case of three
bosons. In the three-particle system the angle $\xi$, determined by the ratio of the
two- and three-body binding energies $E_3/E_2=\tan^2\xi$,
  characterizes the Discrete Scale Invariance of the system.
Extending the definition of the angle to the $N$-body system as $E_N/E_2=\tan^2\xi$,
we study the $N$-boson spectrum in terms of this variable.
The analysis of the results, obtained for
up to $N=16$ bosons, allows us to extract a general formula for
the energy levels of the system close to the unitary limit. Interestingly, a linear dependence of the universal function as a 
function of $N$  is observed at fixed values of $a$. We show that
the finite-range nature of the calculations results in the  range corrections that
generate a shift of the linear relation between
the scattering length $a$ and a particular form of the universal function.
We also comment on the limits of applicability of
the universal relations.

\end{abstract}
\maketitle

\section{Introduction.}

Physical systems present universal behavior when
specific details of the interaction between their constituents are suppressed
in favor of a few control parameters which determine the dynamics. Well-known examples 
of this kind are critical phenomena in which the systems, that are very different at the microscopic level,
 show a set of equal critical exponents.
Around the critical point the dynamics is governed by long range
correlations and not by the details of the interaction
between the constituents. When a universal class is identified
all systems belonging to this class can be described equally well
by a model in which this particular phenomenon is implemented. 
As an example we can mentioned the Ising model to study 
phase transitions. 

Here we analyze a particular universal behavior 
of few-boson systems having a large two-body scattering length.
In two-body systems, the universality  is governed by one parameter, the
scattering length $a$. When $a$ is large and positive the two-body system has a shallow bound state with an
energy of $E_D\approx\hbar^2/ma^2$ (shallow dimer) and, in addition, all the low-energy observables
are governed by $a$. The system has
a continuous scale invariance which strongly constrain the form of the
observables. This symmetry is broken in the $s$-wave three-body sector.
However, the three-body system still has a residual symmetry  - the discrete scale invariance (DSI) - meaning that the physics is invariant under the rescaling $r\rightarrow
\Lambda^n r$, where the constant $\Lambda$ is usually written as
$\Lambda=\text{e}^{\pi/s_0}$, with $s_0\approx 1.00624$ being a universal number
that characterizes a system of three-identical bosons (for a recent review see
Ref.~\cite{braaten:2006_physicsreports})

As has been shown by  V. Efimov~\cite{efimov:1970_phys.lett.b,efimov:1971_sov.j.nucl.phys.},
in the limit of large scattering length, $a\rightarrow\infty$ (unitary limit), 
the three-boson spectrum consists of
an infinite number of states that accumulate to zero with the ratio between the
energies of two consecutive states being $E_3^{n+1}/E_3^n = \text{e}^{-2\pi/s_0}$.
This is known as the Efimov effect and its characteristics have been the subject
of an intense investigation from both
experimental~\cite{ferlaino1,zaccanti2009,ferlaino2,berninger2011} and 
theoretical~\cite{platter2004,hammer2007,stecher2009,deltuva2011,frederico2011} 
point of view. In recent years the study of
the Efimov effect has been extended to what is now called Efimov physics
and refers to the physics of shallow states. In these states the particles
stay mostly far apart from each other with the consequence that the dynamics is largely
insensitive to the details of the interaction.

In the case of bound states, the three-boson spectrum in the limit of zero-range interaction 
in two-body subsystem (zero-range limit or scaling limit) can be expressed
in the following parametric form
\begin{equation}
  \begin{gathered}
    E_3^n/(\hbar^2/m a^2) = \tan^2\xi \\
    \kappa_*a = \text{e}^{(n-n^*)\pi/s_0} 
    \frac{\text{e}^{-\Delta(\xi)/2s_0}}{\cos\xi}\,,
  \end{gathered}
  \label{eq:energyzr}
\end{equation}
where $\kappa_*$ is the wave number corresponding to the energy
of the $n^*$-level at the unitary limit, and is called the three-body
parameter. The function $\Delta(\xi)$ is a universal function and its  
parametrization in the range $[-\pi,-\pi/4]$ is given in 
Ref.~\cite{braaten:2006_physicsreports}. The second line of the above equation explicitly
manifests DSI: the ratio $E_3^{n+1}/E_3^n$ remains constant at each value of the
angle $\xi$ and is equal to $\text{e}^{-2\pi/s_0}\approx 1/515.03$. Another
characteristic of the above equation is that the three-boson spectrum is
controlled by the two-body scattering length $a$ and it is completely 
determined by the knowledge of the three-body
parameter $\kappa_*$, which appears as a scale parameter.  Many examples of 
how Eq.(\ref{eq:energyzr}) is used   can be found in the
literature (see for example Refs.~\cite{braaten:2006_physicsreports,hammer2010}).

In the four-body case, it has been shown that two levels,
$E_4^{n,0}$ and $E_4^{n,1}$ appear attached to each $E_3^n$ level 
\cite{hammer2007,stecher2009,deltuva2011, 
frederico2011,platter2008,hadizadeh2011}.  Moreover, at the unitary limit 
the  ratios  $E_4^{n,0}/E_3^n\approx 4.611$ and 
$E_4^{n,1}/E_3^n\approx 1.002$ are universal and their numerical values have been estimated using different approaches.
In Ref.~\cite{stecher2010} it has been  shown that universal ratios also exist at the universal limit in systems with $N\leq 13$ bosons and these ratios were estimated
by solving the corresponding Schr\"odinger equation with finite-range two-body potentials. 
In Ref.~\cite{gatto2012} the spectra and universality of
small helium clusters up to $N=6$ have been studied in the 
$(1/a,\kappa)$ plane, where 
$\kappa={\rm sign}(E)[|E|/(\hbar^2/m)]^{1/2}$ and $E$ is the energy of a level.
The  two lowest  levels with the energies of
$E_N^{0,0}$ and $E_N^{0,1}$, 
attached to the ground state of the $N-1$ system, with energy of $E_{N-1}^{0,0}$, 
have been observed in the region of $a$ studied. 
Furthermore in Refs.~\cite{stecher2010,gatto2012} the values of $a$ at which the $N$-body
cluster disappears into the $N$-body continuum have been estimated. 

In the present work we address the constraints imposed by the DSI in the spectrum of the $N$-boson system
with  a general number of bosons $N$. In addition we will answer the following questions.
 1)~Is the tree structure of two states, $E_N^{0,0}$ and $E_N^{0,1}$, attached to the
$E_{N-1}^{0,0}$ state and observed up to $N=6$, valid for general values of $N$?
2)~Are there any general relations for the ratios $E_N^{n,0}/E_{N-1}^{n,0}$ and
$E_N^{n,1}/E_{N-1}^{n,0}$? We answer these question by analyzing the spectra of
$N \leq 16$ particles obtained from solution of the many-body Schr\"odinger equation with 
a soft finite-range two-body force.

The paper is organized as follows. In the next section the working 
equations determining the spectrum of the $N$-boson system are given.
The analysis of the numerical results is given in Section III,
whereas the DSI for $N$ bosons is analyzed in Section IV. In section
V the equations proposed in the previous sections are used to analyze 
selected experimental as well as theoretical results from the literature.
In the last section the conclusions are given.

\section{The $N$-boson system in the zero-range limit.} 

Our aim is to discuss an extension of Eq.(\ref{eq:energyzr}), that describes the energy spectrum of the three-boson system close to the unitary limit in the zero-range limit, to $N>3$. 
This extension is based on the detailed  analysis of a four-boson system with a large two-body scattering length, reported first in Ref.~\cite{hammer2007} and then in Refs.~\cite{stecher2009,deltuva2011}, and on an extended analysis of DSI in 
Ref.~\cite{gattoFB2013}.
It has been found  in Ref.~\cite{hammer2007} that the four-body system has
two bound states, one of which is deeply bound and another one is shallow being very close to the threshold of disintegration into one boson and a trimer. Using a DSI argument,
it has been conjectured that the two-level structure is tied to each
three-body state. However, only the lowest two states, attached to the
trimer bound state $E_3^0$, are true bound states;
the other ones appear as resonances since they are above the
trimer-particle threshold. The study of this particular tree structure has  been 
analyzed in Ref.~\cite{stecher2009} in which the notation $E^{n,m}_4$ was
proposed to identify the energy of each four-body level. In this notation, 
$n$ indicates a
three-body level and $m=0$ identifies the deep state while $m=1$ labels the shallow
state. The fact that this structure of levels results from a DSI
can be seen from the universal character of the ratio $E^{n,m}_4/E^n_3$.
A study of this ratio at the unitary limit has been done in 
Refs.~\cite{stecher2009,deltuva2011} with the conclusion that
$E^{n,0}_4/E^0_3\approx 4.611$ and $E^{n,1}_4/E^0_3\approx 1.002$. A more extended
analysis of the DSI can be done studying these ratios along the $(1/a,\kappa)$ plane
at fixed values of the angle $\xi$ (see Ref.~\cite{gattoFB2013}). Moreover,  
a recent work~\cite{gatto2013}  has   shown that the
universal function $\Delta(\xi)$, that governs the three-boson dynamics, 
is also responsible for the $N$-boson dynamics. All these findings suggest the
following extension of Eq.(\ref{eq:energyzr}) to $N\ge 4$ 
\begin{equation}
  \begin{gathered}
    E_N^{n,m}/(\hbar^2/ma^2)= \tan^2\xi \\
    \kappa^{m}_N a  =  \text{e}^{(n-n^*)\pi/s_0}
    \frac{\text{e}^{-\Delta(\xi)/2s_0}}{\cos\xi}\,,
  \end{gathered}
  \label{eq:energygN1}
\end{equation}
with $E_N^{n,m}$ the deep ($m=0$) or shallow ($m=1$) $N$-body state 
attached to the $n$-th Efimov trimer.

Eq.~(\ref{eq:energygN1}) has the remarkable property that the $N$- boson
spectrum is controlled by the two-body scattering length, by the
universal function $\Delta(\xi)$, and that
it is completely determined by the knowledge of $\kappa^{m}_N$. 

The $\kappa^{m}_N$ are not true-independent parameters; they are
fixed by the three-body scale $\kappa_*$. For instance, in Ref.~\cite{gatto2013}
it has been shown that 
\begin{equation}
\kappa_N^0/\kappa_*=1+1.147(N-3)\,.
  \label{}
\end{equation}
This result has been obtained by observing that $\kappa^{m}_N$
is  a linear function of $N$ and using the
universal ratio given in Ref.~\cite{deltuva2011}. In the following
we analyze the spectrum of $N$-boson systems up to $N=16$ in order  
to extend this relation to $m=1$ and, in a more general perspective,
to verify the validity of Eq. (\ref{eq:energygN1}).

\section{Finite-range corrections}

The spectrum of the three-boson system, obtained by solving the Schr\"odinger equation with soft two-body potentials close to the unitary limit, has been analyzed in Refs.~\cite{kievsky2013,garrido2013}. It has been shown  that Eq.(\ref{eq:energyzr}) has to be modified in order to take into account the finite-range character of those calculations.
Based on the  analysis of the energies of the three-body system obtained numerically, the following modified equation  has been deduced 
\begin{equation}
  \begin{gathered}
    E_3^n/E_2= \tan^2\xi \\
    \kappa^n_3 a_B + \Gamma^n_3 = \frac{\text{e}^{-\Delta(\xi)/2s_0}}{\cos\xi}\,.
  \end{gathered}
  \label{eq:energyN3}
\end{equation}
Despite some similarities, there are several important differences to  the zero-range theory of Eq.(\ref{eq:energyzr}). 

(i) The parameters $\kappa^n_3$
carry explicitly the index $n$ labeling the different tree  branches since
the ratio $\kappa^n_3/\kappa^{n+1}_3$ for two successive branches  is in general 
slightly different from the scaling factor $\text{e}^{\pi/s_0}$. In fact these ratios
include finite-range corrections and their specific values can be extracted from
the numerical solutions.
The correspondence between Eq.(\ref{eq:energyzr}) and Eq.(\ref{eq:energyN3}) is
made by identifying the scale parameter $\kappa_*$ with one of the parameters
$\kappa^n_3$. For example in Ref.~\cite{kievsky2013} the three-helium atom case
was studied as a reference system and $\kappa_*$ has been
identified with $\kappa^1_3$, the branch corresponding to the first excited state. 

(ii) The quantity $\hbar^2/m a^2$ is replaced by
$E_2=\hbar^2/m a_B^2$, which is the dimer binding energy in the case of positive 
scattering length $a$ or, for negative values of $a$, the two-body virtual-state energy.
Also in the second line $a$ is replaced by the scattering length $a_B$ corresponding to   finite-range two-body potential. The replacement of $a$
by $a_B$ introduces
some range corrections as the value of $a$ moves away from the unitary limit.
It should be noticed that the relation $E_2=\hbar^2/m a^2$ is exact 
in the case of zero-range two-body interactions. 

(iii) The main modification in Eq.(\ref{eq:energyN3})
is the introduction of the shift parameter $\Gamma^n_3$.
 The origin of the shift has been discussed in Ref.~\cite{gatto2013} where
it has been shown that it essentially appears from the first order expansion  
of the scaling-violating momentum $\Lambda_0$,  in terms of powers of $r_0/a$, 
where $r_0$ is the interaction range.  
The parameter $\Lambda_0$ fixes the value of the logarithmic
derivative of the wave function close to the origin and encodes the short-range
physics \cite{braaten:2006_physicsreports}.
In the zero-range model one can identify $\Lambda_0$ with the three-body
parameter $\kappa_*$ 
leading directly to Eq.(\ref{eq:energyzr}). However, for finite-range potentials
the $\Lambda_0=\kappa_*$  relation does not hold anymore and we propose the finite-range correction
\begin{equation}
\Lambda_0=\kappa_* (1+{\cal A}\frac{r_0}{a}+\ldots) \, ,
\end{equation}
leading directly to Eq.(\ref{eq:energyN3}), with 
$\Gamma^0_3={\cal A}\kappa^0_3 r_0$ (here we assume $\kappa_*=\kappa^0_3$). This can be  
proved by making use of Eqs.(187) and (193) from  Ref. \cite{braaten:2006_physicsreports}.

As a function of the finite-range-corrected scattering length $a_B$,
Eq.(\ref{eq:energyN3}) is a two-parameter equation: the scale parameter
$\kappa^n_3$ and the finite-range parameter $\Gamma^n_3$.  As in the study of
critical phenomena, $\kappa^n_3$ can be interpreted as a material-dependent
parameter, and  $\Gamma^n_3$ as the
analogue of the finite-scale correction (here finite-range correction). Their
introduction allows the collapse of observables onto a single universal
curve~\cite{stanley:1999_rev.mod.phys.}.  In fact, in Ref.~\cite{gatto2013}, it
has been shown that  plotting $E^n_3/E_2$ in terms of
$1/(\kappa^n_3a_B+\Gamma^n_3)$ makes the calculated points to collapse onto a universal
curve. 

The extension of Eq.(\ref{eq:energyN3}) to general $N$ has been proposed in
Ref.~\cite{gatto2013} where the spectrum of the $N$-boson system,
obtained by solving the Schr\"odinger equation with soft potentials 
close to the unitary limit, has been analyzed up to $N=6$.
The results of that work have shown that Eq.(\ref{eq:energygN1}) has to be
modified in order to take into account the finite-range character of those
calculations. The following general form has been deduced 
\begin{equation}
  \begin{gathered}
    E_N^{n,m}/E_2= \tan^2\xi \\
    \kappa^{n,m}_N a_B + \Gamma^{n,m}_N = \frac{\text{e}^{-\Delta(\xi)/2s_0}}{\cos\xi}\,,
  \end{gathered}
  \label{eq:energygN}
\end{equation}
and applied to the lowest tree of the two-level structure which correspond to 
$n=0$. The index $n$ identify the $N=3$ branch and the index $m$ takes the value $m=0,1$.
However, unlike in the zero-range theory, the parameters $\kappa^{n,m}_N$
explicitly show the index $n$ labeling the different $N=3$ branches due to the
finite-range corrections in $\kappa^n_3$. 

It should be stressed that, for 
$N>3$, the $n=0$ case is of particular interest since it describes bound states.
In the next section, we analyze the
$n=0$ branch extracting the $\kappa^{0,m}_N$ values from the numerical solutions and showing 
that they have a linear dependence with $N$ with a slope slightly different 
to the one suggested by the zero-range theory. We will show that, as $a$ moves
from the unitary limit toward lower positive values, the validity of
Eq.(\ref{eq:energygN}) could be limited. The system becomes more compact losing its 
universal character. In the case of $n>0$ the validity of Eq.(\ref{eq:energygN})
is limited by the appearance of different thresholds as the positive values of
$a$ decrease. For example in Ref.~\cite{deltuva2013} it was shown that, 
in the $N=4$ case, a shallow tetramer decays at the atom-trimer threshold becoming 
an inelastic virtual state. In the present work we limit the discussion of
Eq.(\ref{eq:energygN}), in the case of $N>3$, to the case of bound states ($n=0$ branch).

\section{Analysis of $N$-body solutions.}

To study the validity of Eqs.(\ref{eq:energygN1}) and 
(\ref{eq:energygN}) we
follow  Ref.~\cite{gatto2012} and
describe the $N$-boson system using a two-body gaussian (TBG) potential
\begin{equation}
 V(r)=V_0 {\rm e}^{-r^2/r_0^2}\,.
\end{equation}
We solve the $N$-body Schr\"odinger equation with mass parameter $\hbar^2/m=43.281307$
$(a_0)^2$K. Using $r_0=10\, a_0$ and $V_0=-1.2343566\,$K the model reproduces 
the binding energy and the scattering length of two helium atoms described
by a  widely used He-He interactions, the LM2M2
potential~\cite{lm2m2} which has a van der Waals length $\ell=10.2\,$ a.u..

To solve the Schr\"odinger equation for $N$ bosons we use the Hyperspherical
Harmonic (HH) method in the version proposed in Ref.\cite{natasha1}. This method
reproduces the values given in Ref.~\cite{gatto2011} up to $N=6$ and here we
extend the calculations up to $N=16$. Increasing the grand angular quantum 
number $G$ we obtain converged results for the ground state and first excited state
of the $N$-boson systems. As discussed in
Refs.~\cite{natasha1,natasha2},
convergence of the ground state energy is obtained with relatively low
values of $G$; with values of $G\le12$ an accuracy greater than 1\% is
obtained. In the case of the first excited state a similar accuracy 
would need a much higher value of $G$, making the computation of this state
very difficult. In the present work we use the results with $G\le 12$ to extrapolate 
the first excited energy with an accuracy of a few percent.

Varying the strength $V_0$ of the TBG potential (7) we explore the
$(a^{-1},\kappa)$ plane.
For each value  of the potential strength $V_0$  we determine $E_2$ and $a$
and then compute the energies of
the ground and first excited states of the $N$-body systems. For $N=3$ we compute $E_3^0$ and $E_3^1$ - the  values that define the first two
energy branches with $n=0$ and $n=1$. For $N>3$ we
compute the two-level structure of the $n=0$ branch. We denote these states as
$E_N^0$ and $E_N^1$, omitting the index $n$ from now on, and we analyze 
the two-level spectrum up to $N=16$.

Using the first line of Eqs.~(\ref{eq:energyN3}) and (\ref{eq:energygN})
we determine the value of the angle $\xi$ from which we can compute  
the universal function $y(\xi)$
\begin{equation}
    y(\xi)= \frac{\text{e}^{-\Delta(\xi)/2s_0}}{\cos\xi}\ .
\end{equation}
appearing on the r.h.s of the second line of these equations. The function
$y(\xi)$ has a linear dependence on $a_B$ for each value of $N$ and $m$. 
In fact, the second line of Eq.(\ref{eq:energygN}) can be rewritten as
\begin{equation}
  y=\kappa^m_N a_B+\Gamma^m_N\,.
    \label{eq:linear}
\end{equation}
In the case of zero-range interaction $a_B=a$  and $\Gamma_N^m=0$,
and the linear relation results in $y=\kappa_N^m a$, representing a straight 
line passing through the origin with the slope determined by the value of $\kappa_N^m$. 
In the case of finite-range interaction the linear dependence between $y$ and
$a_B$ remains but the straight line does not go through the origin. 

\begin{center}
  \begin{figure}[t]
    \includegraphics[width=10.0cm,angle=-90]{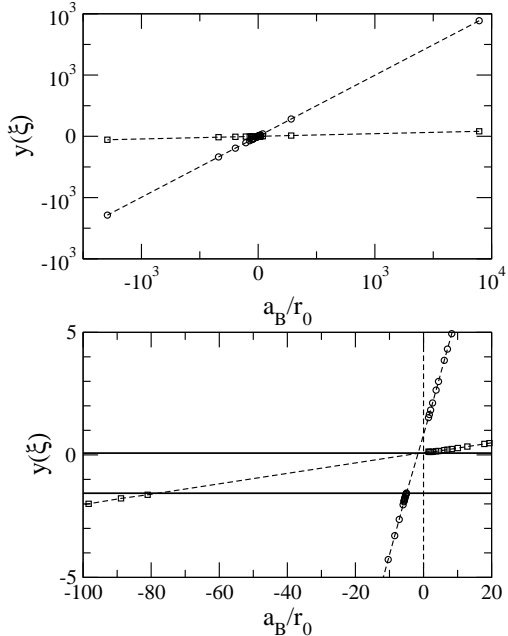}
    \caption{The universal function $y(\xi)$ as a function of $a_B$ in units
of $r_0$. Numerical results are given by circles 
($n=0$) and squares ($n=1$) with the dashed lines representing the best linear
fits to the results (upper panel). In the lower panel a zoom close to the
thresholds $y(-\pi)=-1.56$ and $y(-\pi/4)=0.071$ (given by the thick horizontal
lines) is shown. 
}
  \label{fig:k3star}
  \end{figure}
\end{center}

To make connection with previous analysis, we first present our results for
the $N=3$ case.
Their plot in  the $(a^{-1}-\kappa)$ plane have been reported many times in the literature. Here we prefer to 
use the $(a_B-y)$ plane stressing the linear relation between $a_B$ and $y$ that follows 
from  Eq. (\ref{eq:linear}). The results are shown in Fig.~\ref{fig:k3star} as circles
($n=0$ branch) and squares ($n=1$ branch). 
The dashed lines represent the
best linear fit to the results which can be parametrized with $\kappa_3^0=0.0488$
a.u.$^{-1}$, $\Gamma_3^0=0.869$ and $\kappa_3^1=0.00212$ 
a.u.$^{-1}$, $\Gamma_3^1=0.0840$. The fit has $\chi^2\approx 0.1$ proving that the
behavior of the numerical results is in very good agreement with a linear dependence. Moreover, at the unitary limit the 
extracted values for $\kappa_3^n$ coincide up to four figures (or better) to
the calculated values. 
The DSI  predicts in the zero-range limit that $\kappa_3^0/\kappa_3^1=e^{\pi/s_0}\approx 22.7$
while from our
results we obtain a slightly different value, $\kappa_3^0/\kappa_3^1\approx 23.0$. This change
originates  due to the
finite-range character of the two-body interaction. 

The upper panel of Fig.~\ref{fig:k3star} shows all calculated $y(\xi)$ values plotted within a
very extended range of $a_B$. The lower panel displays a zoom around  the
thresholds $y(-\pi)\approx-1.56$, at which  the trimer levels  disappear
into the three-body continuum (for a more precise value see Ref.~\cite{gogolin}), 
and $y(-\pi/4)\approx 0.071$, at which the trimer
disappears on the particle-dimer continuum (the thresholds are shown by thick
horizontal lines). Close 
to the threshold at -1.56 we observe a strong linear trend in the $y(\xi)$ behavior
as a function of $a_B$. This allows us to extract
with a great confidence the (negative) values of the scattering length $a^{0,-}_3$ and
$a^{1,-}_3$ at which the trimer ground state $E_3^0$ and first excited
$E_3^1$ disappear into the three-atom continuum. 
Using the values of $\kappa_3^n$ and $\Gamma_3^n$ from the linear fit and
transforming $a_B$ to $a$, we obtain
$a^{0,-}_3\approx -44\,$a.u. and $a^{1,-}_3\approx -745\,$a.u.  which are in close
agreement with the estimates given in Ref.~\cite{gatto2012}.


The threshold at 0.071 indicates the point at which the bound state disappears
in the atom-dimer continuum. We cannot  reach it by lowering the scattering length  
because  as $a\rightarrow r_0$ the three-body states become more bound leaving
the Efimov window. For the case of the excited state this is less obvious but
in the  proximity to the atom-dimer threshold our 
results 
lie on a line parallel to (and slightly above) the $y(-\pi/4)$ line without
crossing it. This has been also clearly seen in Ref.~\cite{gatto2012} where
the first excited state for three Helium atoms has been shown  not to cross
the atom-dimer threshold but move almost parallel to it from below
(see also Ref.~\cite{naidon}).

\begin{figure*}[t]
    \includegraphics[width=14cm,angle=-90]{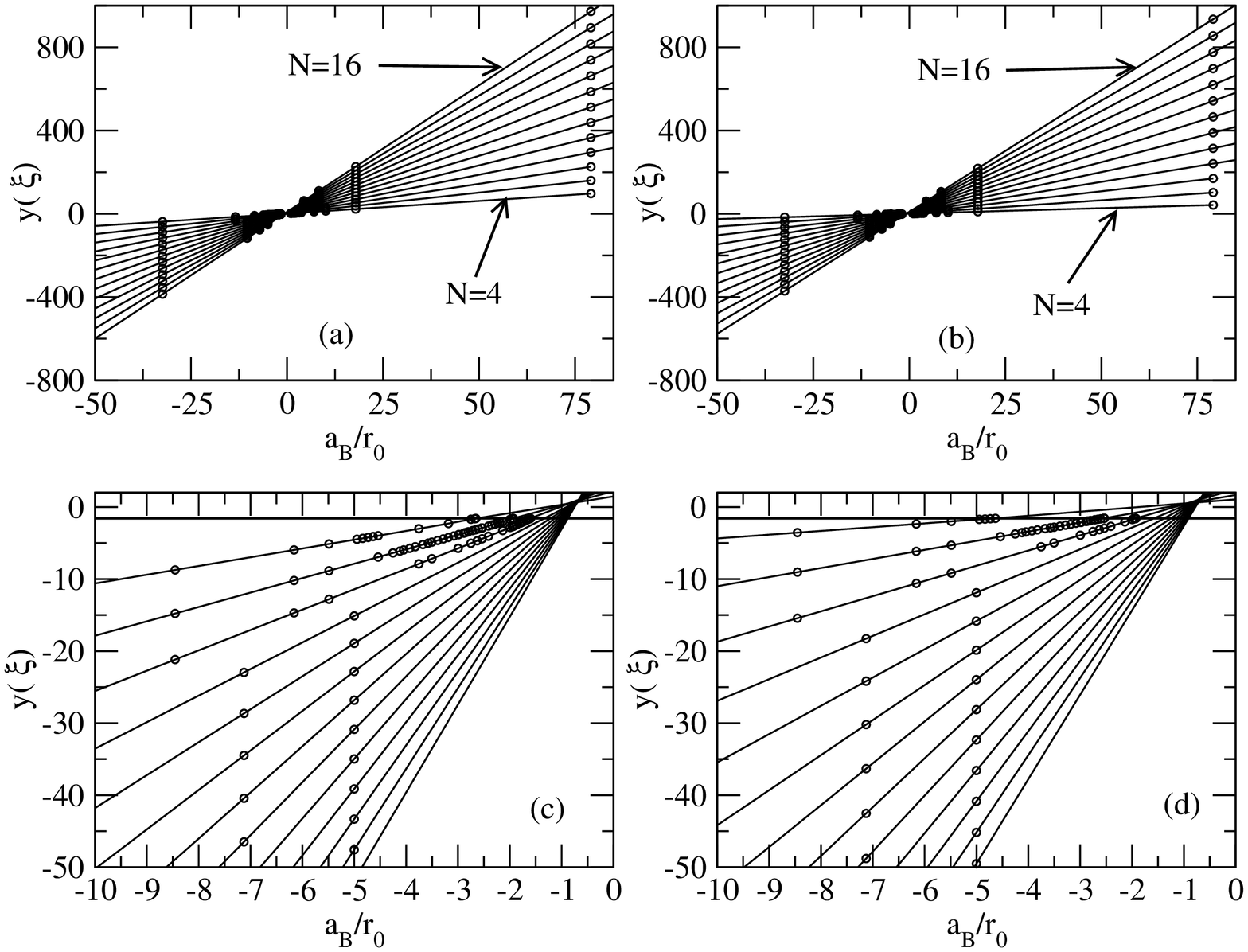}
    \caption{The universal function $y(\xi)$ as a function of $a_B$,
in units of $r_0$ for $N=4-16$. (a)Ground state energies and (b) 
excited state energies. A zoom of the plots close to the -1.56
threshold is given in (c) for $n=0$ and in (d) for $n=1$}
  \label{fig:linear}
 \end{figure*}

The
analysis of the Efimov trimers in the $(a_B-y)$ plane has shown that they can be
described by straight lines defined by parameters
$\kappa_3^n$, $\Gamma_3^n$ and $a^{n,-}_3$. 
Next,  we study the   $N\ge 4$ results for $y(\xi)$ in the same $(a_B-y)$ plane.
These results are presented by circles in Fig.~\ref{fig:linear} which, as in the
$N=3$ case,
can be fitted by straight  lines shown as solid lines in this figure. 
Panels (a) and (b) show the results for the ground state ($m=0$) and the first excited state  ($m=1$) respectively. Panels (c) and (d) 
  zoom the area around the threshold $y(-\pi)=-1.56$ (shown as a thick line). 
In the case of $N=4,5,6$ very 
detailed calculations has been done close to this threshold.

\begin{center}
  \begin{figure}[t]
    \includegraphics[width=12cm,angle=-90]{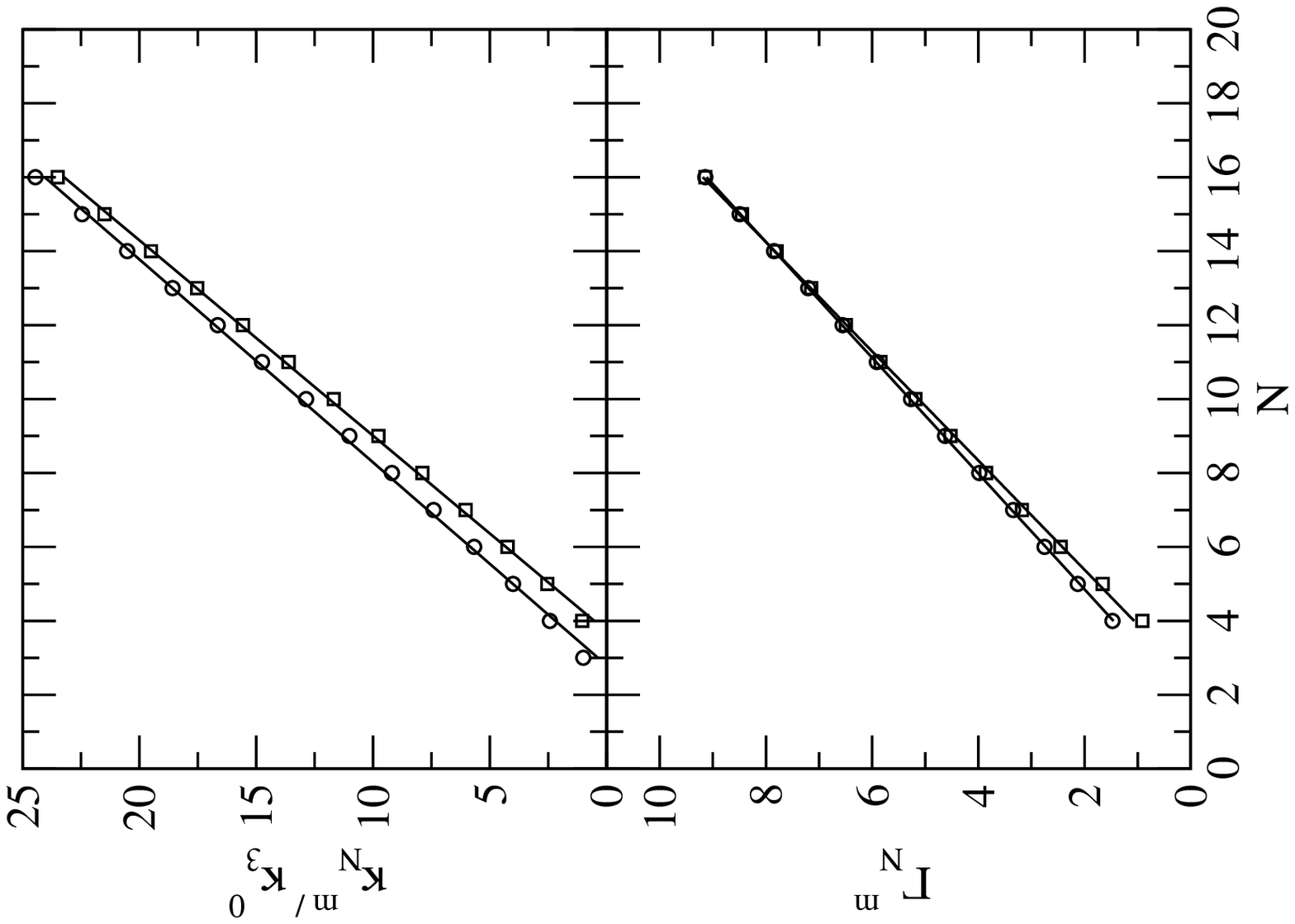}
    \caption{The  ratio $\kappa_N^m/\kappa_3^0$ (upper panel)
and the  shift $\Gamma_N^m$ (lower panel) as a function of $N$. 
The circles correspond to $m=0$ and the squares to $m=1$. The solid lines are
the best linear fits.}
  \label{fig:gamma1}
  \end{figure}
\end{center}

\begin{center}
  \begin{figure}[t]
    \includegraphics[width=6cm,angle=-90]{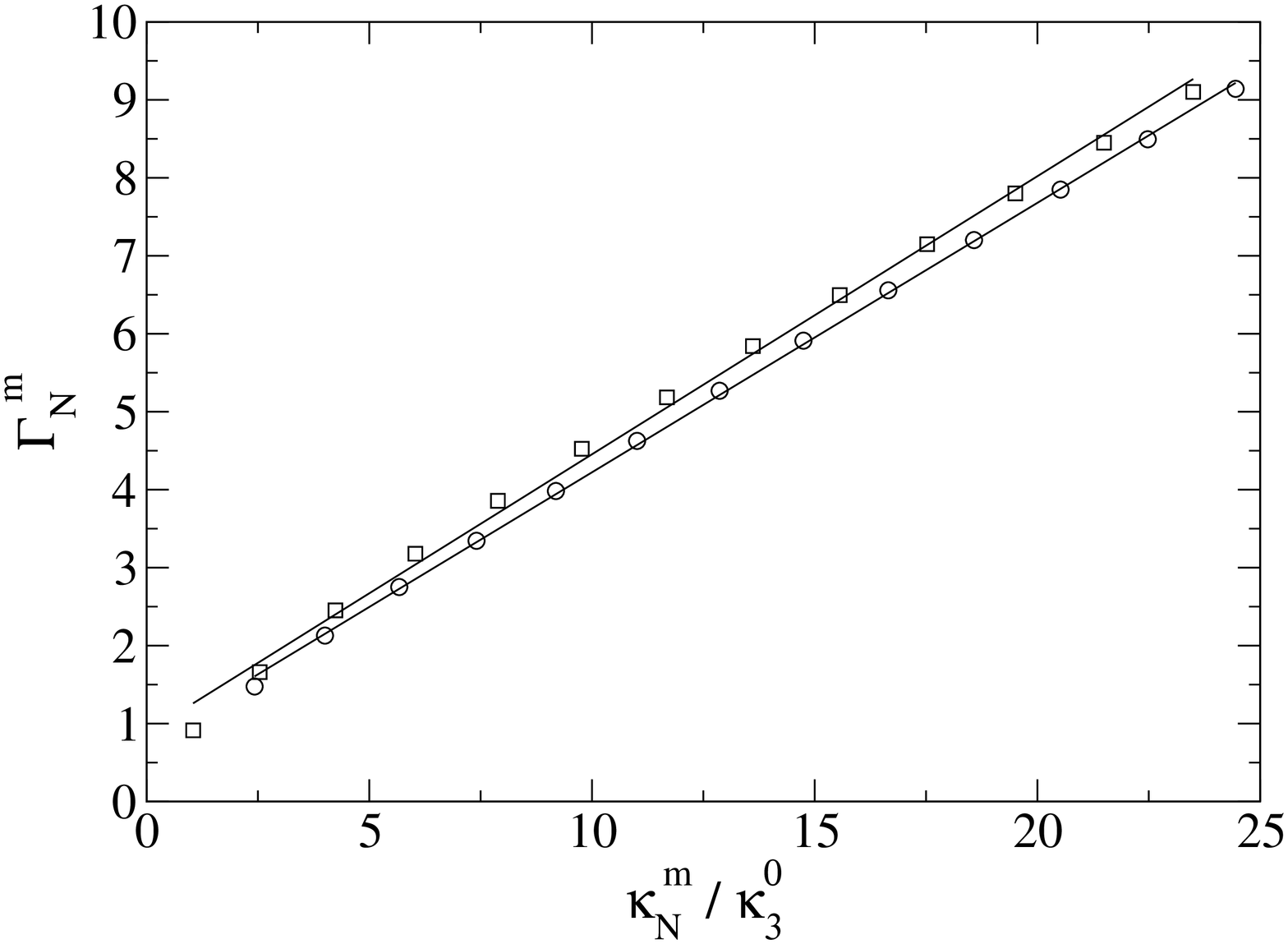}
    \caption{The shift $\Gamma_N^m$ as a function of the ratio
$\kappa_N^m/\kappa_3^0$ for $m=0$ (circles) and $m=1$ (squares).}
  \label{fig:onepoint}
  \end{figure}
\end{center}

\begin{figure}[t]
    \includegraphics[width=7cm,angle=-90]{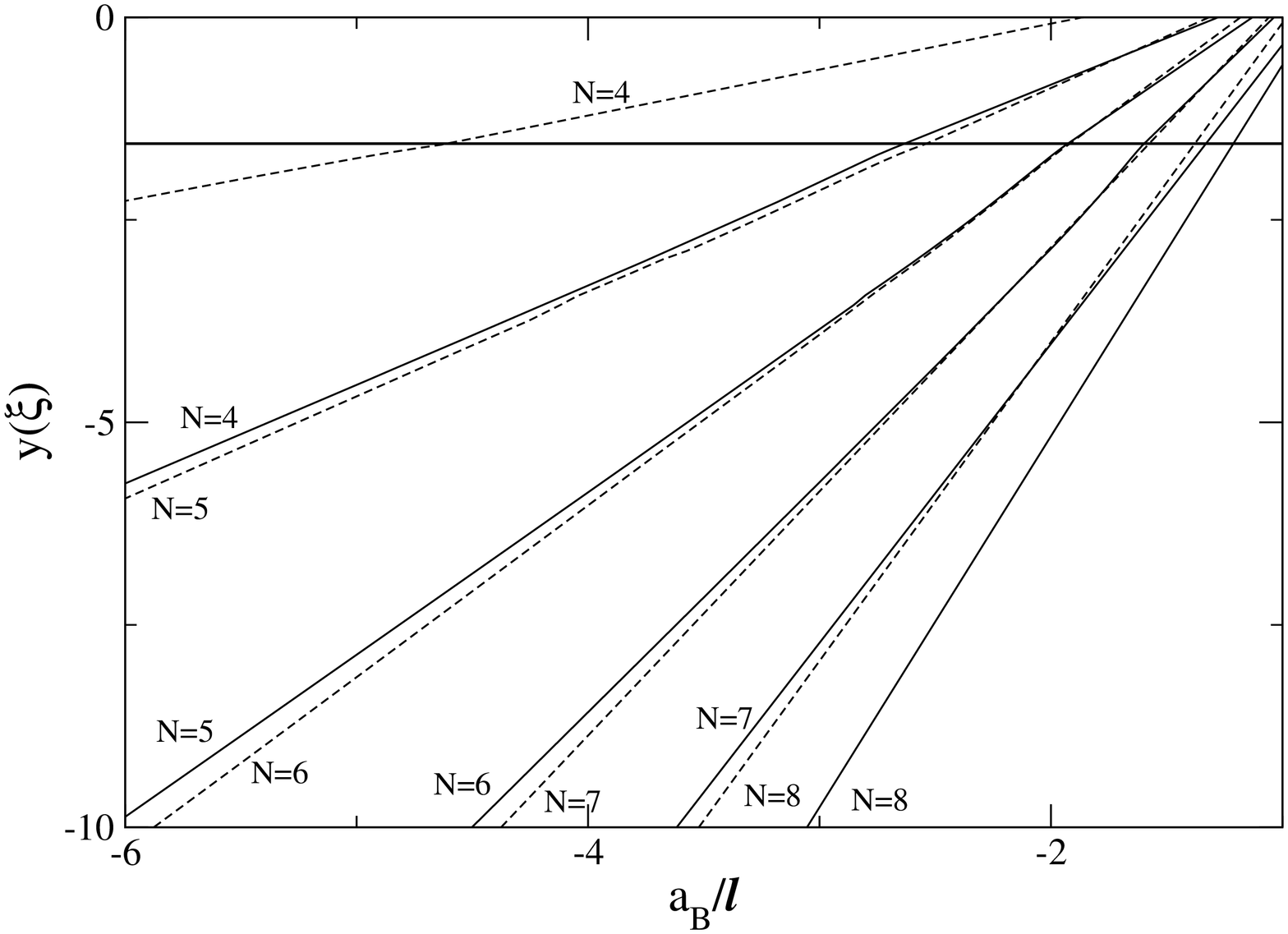}
    \caption{The universal function $y(\xi)$ as a function of
$a_B$, in units of $\ell$ for the ground state (solid lines)
and excited state (dashed lines) close
to the $y(-\pi)=-1.56$ threshold, for $N=4-8$. Starting with $N=8$ the
excited state is becomes unbound at this threshold as it crosses
the ground state line before arriving to the threshold.}
  \label{fig:crossing}
 \end{figure}

A remarkable characteristic of Fig.~\ref{fig:linear} is that the straight lines 
 cross each other almost in one single
point the position of which is slightly different for the ground and   excited states.
To study further this fact we extract the values
of $\kappa_N^m$ from the slopes of the straight lines. They coincide up to four
significant figures with the values calculated at the unitary limit given by
$V_0\approx -1.162$ K. The results are given in Fig.~\ref{fig:gamma1} where we
show: (a) the values of $\kappa_N^0$ (circles) and $\kappa_N^1$ (squares)
 as a function of $N$ and (b) the values of $\Gamma_N^0$ (circles) and 
$\Gamma_N^1$ (squares). The  $\kappa_N^0$ and $\kappa_N^1$ form two almost
parallel lines whereas  $\Gamma_N^0$ and $\Gamma_N^1$  collapse in one line
as $N$ increases. However,   both   $m=0$ and $m=1$ explicitly show a linear dependence
with $N$,  which is illustrated in Fig. ~\ref{fig:onepoint} where
the best linear fits to the data are shown
by the solid lines and the circles (squares)
corresponds to $m=0$ ($m=1$).  Using this linear relation we extract
the coordinates of the point $(a_B^m,\Gamma_m)$ at which the straight lines 
corresponding to the ground states, $m=0$, and excited state, $m=1$, cross
each other. Defining $\Gamma_N^m=\Gamma_m-\kappa_N^m a_B^m$, we obtain the relation between
$\kappa_N^m$ and $y$:
\begin{equation}
\kappa_N^m (a_B-a_B^m) +\Gamma_m=y(\xi)
\label{eq:linear1}
\end{equation}
From the analysis of Fig.~\ref{fig:onepoint} we get
$a_B^0=7.077$ a.u. and $\Gamma_0=0.768$ for the ground states
and $a_B^1=7.304$ a.u. and $\Gamma_1=0.887$ for the excited states. 
A consequence of the different location of these two points is
that the line corresponding to the shallow state $E_N^1$ can cross
the line of the ground state $E_{N-1}^0$ resulting in unbound excited state. This is
shown in Fig.~\ref{fig:crossing} where one can see that
starting with $N=8$ the  $E_N^1$ excited state is not bound anymore 
at the $y(-\pi)$ threshold.

Eq. (\ref{eq:linear1}) can be further simplified taking into account the
linear relation between $\kappa_N^m$ and $N$. In fact, as has been
inferred in Ref.~\cite{gatto2013}, and from the analysis of 
Fig.~\ref{fig:gamma1}, the relation can take the following form,
\begin{equation}
\kappa_N^m= \kappa_4^m +(N-4)(\kappa_5^m-\kappa_4^m),
\label{eq:linear2}
\end{equation}
where we have used two points, corresponding to $N=4$ and $N=5$ to construct
the line. However other choices to describe the straight line produces similar results. 
With the above relation, a global description of the $m$th level of the 
$N$-boson spectrum can be achieved with just four parameters.

The main result of this analysis is the following. In order to describe the
$m$th energy level of the $N$-boson system, using short range interactions, 
two parameters are needed, $\kappa_N^m$ and $\Gamma_N^m$. This description is
particularly
accurate near   the unitary limit. It deteriorates at  positive values of $a$ as $a\rightarrow \ell $ where the strength of 
the potential increases making the $N$-body  ground states deeper. 
In addition, as $N$ increases the finite character of the interaction allows
new excited states to appear and the tree structure is lost. It was shown in
Ref.~\cite{cracow} that starting with $N\ge12$ a second excited state appears at the unitary
limit. However, the two-parameter
description remains acceptable (of the order of a few percent or better)
as the system approaches the $N$-body continuum (negative $a$ values),
which corresponds to the best realization of shallow states, well described by
the present formalism. Approaching the threshold on the $-\pi$ axis
all the excited states disappear for $N>7$, as has been discussed before.
It should be noticed that in the case of a zero-range interaction there is only one parameter,$\kappa_N^m$, because $\Gamma_N^m=0$,  and, as
the crossing point is the origin for the two $m$-levels, the tree structure
should remain valid with increasing $N$. 

In the case of finite-range interactions, using the relation given by 
Eq.(\ref{eq:linear2}),
a global fit of the $m$th level is possible for general values of $N$
with only four parameters: two $\kappa_N^m$ values and the
coordinates of the crossing point. However, this description is not as precise
as the previous one and could introduce some errors.
In the case of a zero-range interaction a global fit is possible
with two parameters (two values of $\kappa_N^m$) and,
in this case the description should be exact. This is further analyzed in the
next section.

\section{The DSI as a function of $N$.} 

The DSI in the $N=3$ system can be seen
from the constant values of the ratio $E_3^n/E_3^{n'}$ between two different
branches at fixed values of the angle $\xi$.
This property, in the zero-range theory, is encoded in Eq.(\ref{eq:energyzr}) 
from which it is easy to see that, when $\xi$ is constant, then
$E_3^n/E_3^{n'}=\text{e}^{2(n'-n)\pi/s_0}$. Particular cases are the unitary 
limit corresponding to $\xi=-\pi/2$ and the threshold at which the cluster
disappear in the three-atom continuum corresponding to $\xi=-\pi$. However
that property holds in all the range $-\pi\le \xi \le -\pi/4$.

The extension of the zero-range theory to general $N$ 
is given in Eq.(\ref{eq:energygN1}). This equation
explicitly states the following property for the ratio
between two different branches with general values of $m$ at fixed values of $\xi$
\begin{equation}
\frac{E_N^{n,m}}{E_{N'}^{n',m'}}= 
\left(\frac{\kappa^m_N}{\kappa^{m'}_{N'}}\right)^2 \text{e}^{2(n-n')\pi/s_0}. 
\label{eq:ratiosN}
\end{equation}
Unlike in the $N=3$ case,   no  analytical
expression for $\kappa^m_N/\kappa^{m'}_{N'}$ exists. It can be only determined from numerical
analysis such as the one carried out  in Ref.~\cite{deltuva2011} for  
$E_4^0/E_3^0$. However, we can show that $\kappa^0_N/\kappa^{0}_{3}$ 
depends linearly on $\kappa^0_4/\kappa^{0}_{3}$ by using Eq.(\ref{eq:linear2}) 
twice, for $N=3$ and $N=4$. This results in formula (see also
Ref.~\cite{gatto2013})
\begin{equation}
    \frac{\kappa^0_N}{\kappa^0_3} = 1+(N-3)(\frac{\kappa_4^0}{\kappa_3^0}-1)
  \label{eq:kappaN}
\end{equation}
which in the zero-range limit reduces to $\kappa^0_N/\kappa^0_3=1+1.147(N-3)$ when $\kappa^0_4/\kappa^0_3=2.147$ from Ref.~\cite{deltuva2011} is used. Our numerical results using the TBG
potential gives $\kappa^0_4/\kappa^0_3=2.42$, showing some range corrections
but not far from the zero-range limit. The square of Eq.(\ref{eq:kappaN})
gives a quadratic dependence on $N$ of the $N$-boson ground state energy 
$E_N^0$ expressed in terms of the three-boson ground state energy. A quadratic relation
in terms of $N$ has been also obtained in Ref.~\cite{nicholson}.

\begin{center}
  \begin{figure}[t]
    \includegraphics[width=6.0cm,angle=-90]{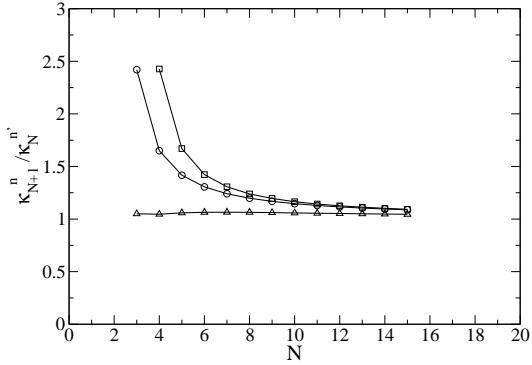}
    \caption{The ratios $\kappa^0_{N+1}/\kappa^0_N$ (circles),
$\kappa^1_{N+1}/\kappa^0_N$ (triangles) and $\kappa^1_{N+1}/\kappa^1_N$
(squares) as a function of $N$.}
  \label{fig:kratios}
  \end{figure}
\end{center}

Our results can be used to
study different ratios that might display universal character.
In Fig.~\ref{fig:kratios} we show the ratios between ground states
$\kappa_{N+1}^0/\kappa_N^0$ (circles), between the shallow state of the $N+1$ system
and the ground state of the $N$-body system, $\kappa_{N+1}^1/\kappa_N^0$ 
(triangles), and the ratio between the shallow states, 
$\kappa_{N+1}^1/\kappa_N^1$ (squares). At large $N$ the ratio 
$\kappa_{N+1}^n/\kappa_N^n$ tends to one suggesting that
\begin{equation}
    \kappa_{N+1}^0/\kappa_N^0\approx \kappa_{N+2}^1/\kappa_{N+1}^1  \,\, .
  \label{eq:kratios}
\end{equation}
This relation is a consequence of the almost constant behavior of the
ratio $\kappa_{N+1}^1/\kappa_N^0$ (see Fig.~\ref{fig:kratios}) that allows  Eq.(\ref{eq:kappaN}) to be extended to shallow excited states of the tree structure for
$N\ge 4$ 
\begin{equation}
    \frac{\kappa^1_N}{\kappa^1_4} = 1+(N-4)(\frac{\kappa_5^1}{\kappa_4^1}-1),
  \label{eq:kappaN1}
\end{equation}
which in the zero-range theory reduces to $\kappa^1_N/\kappa^1_4=1+1.147(N-4)$.
The third  general ratio 
$\kappa_{N+1}^1/\kappa_N^0 =\kappa_4^1/\kappa_3^0$ can be obtained
by observing  that  $\kappa_{N+1}^1/\kappa_N^0$ have a constant behavior. 
For $\kappa_4^1/\kappa_3^0$ the universal ratio of 1.002 can be used, which has been
calculated  in Ref.~\cite{deltuva2011} in the zero-range theory making a detailed 
numerical analysis of the solution of the Faddeev-Yakubovsky equations
(for comparison, our numerical results obtained with TBG potential
give a value around $1.05$). 
This analysis complete the determination of the ratios given in Eq.(\ref{eq:ratiosN}).

\begin{center}
  \begin{figure}[t]
    \includegraphics[width=12.0cm,angle=-90]{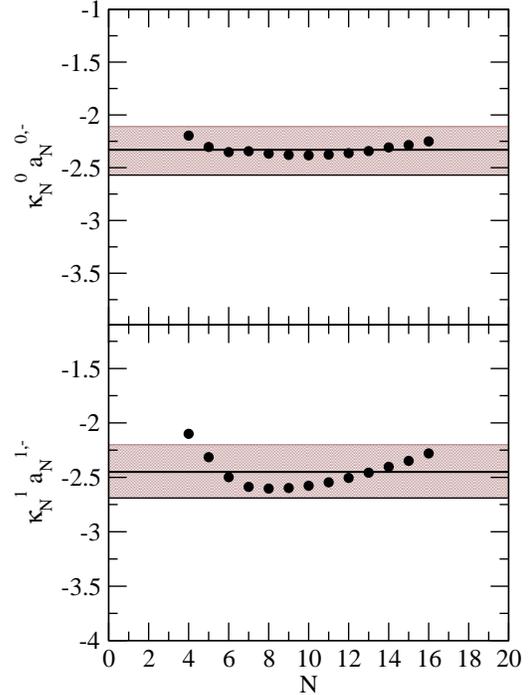}
    \caption{The dimensionless quantity $\kappa_N^na_N^{n,-}$ as
a function of $N$ for $n=0$ (upper panel)and for $n=1$ (lower panel)}
  \label{fig:aminus}
  \end{figure}
\end{center}

Finally, Eq.(\ref{eq:energygN1}) can be used to determine the relation
between $\kappa^m_N$ and $a$ at different thresholds $\xi=-\pi$
at which the $N$-body cluster disappears in the $N$-body continuum.
Using the notation 
$a_N^{m,-}$ for the corresponding value of $a$, in the zero-range
limit   we have
\begin{equation}
    \kappa^m_N a_N^{m,-} = -{\rm e}^{-\Delta(-\pi)/2s_0}\approx -1.56
  \label{eq:aminusN}
\end{equation}
This equation is an extension of the already known relation obtained
in the zero-range $N=3$ case. 
In the case of finite-range interactions we use the working equation  
Eq.(\ref{eq:energygN}) and verify that the following
modification of Eq.(\ref{eq:aminusN}) is valid: 
\begin{equation}
    \kappa^m_N a_N^{m,-} \approx -1.56 -\Gamma_m  \, .
  \label{eq:aminusNn}
\end{equation}
Here $a_N^{m,-}=a_B(-\pi)-a_B^m$, with $a_B(-\pi)$ being the value at the threshold determined
either from direct calculations or from a global  linear fit.
We evaluate  $\kappa^m_N a_N^{m,-}$ using the calculated values for 
$\kappa^m_N$ and $a_N^{m,-}$.
The results are given in Fig.~\ref{fig:aminus} for both $m=0$ and $m=1$. One can
see that in both cases the quantity $\kappa^m_N a_N^{m,-}$ is close to
the expected values of -2.33 ($m=0$) and -2.45 ($m=1$). For $m=0$ 
$\kappa^m_N a_N^{m,-}$ is better represented by a constant
supporting the global fit. For $m=1$ the quantity $\kappa^m_N a_N^{m,-}$
deviates from a constant within 10\% (given by
the shadowed area) since excited states' binding energies have not completely converged.  
In particular the global fit for 
the excited states suffers from the fact that close to the
$y(-\pi)$ threshold they are not  bound any more for $N>7$. We have to mention here that
in Ref.~\cite{stecher2010}  a different
parametrization of $a_N^{0,-}$ exists that involves four parameters. Such a parametrization
does not show universality.

In the case of positive values of $a$,
thresholds appear when the $N$-body systems disappears into the continuum
formed by the different clusterization of the $N$-boson system. The first
threshold appears at the $(N-1)$-boson energy and it is formed by a cluster
of $(N-1)$ bosons and one boson staying far apart. Other thresholds are formed by 
$N/2$ dimers (for even values of $N$) or by $(N-1)/2$ dimers plus
a particle (for odd values of $N$). In the case of $N=4$ a detailed study
of the behavior across the dimer-dimer and trimer-atom thresholds has been done in
Ref.~\cite{deltuva2013}. With increasing   number of bosons the structure of the
thresholds become more and more complicate. As discussed in the $N=3$ case, the
validity of Eq.(\ref{eq:energygN1}) for each $N$ system is limited by values of 
$\xi$ between $\xi=-\pi$ and the appearance of the first threshold. An study of the 
behavior of the $N$-boson system across these thresholds is at present underway.

\section{Analysis of the results in different systems}

Most of the studies of $N$-boson systems with large two-body scattering
lengths have been done either experimentally  (real systems) or solving
the Schr\"odinger equation using potential 
models for two-boson interaction. For example, here we have used a TBG potential,
however, many other calculations with different model or realistic potentials
can be found in the literature. In all these cases 
the deviations from the prediction of zero-range theory due to finite-range
nature of two-body interactions cannot be ignored.
The working equations,  Eqs.(\ref{eq:energyN3}) and (\ref{eq:energygN}),  proposed here
can be used to analyze results obtained elsewhere. In particular,
the second line of these equations can be used to study the linear dependence
between $y$ and $a_B$ (see Eq.(\ref{eq:linear})).

\begin{center}
  \begin{figure}[hb]
    \includegraphics[width=7.0cm,angle=-90]{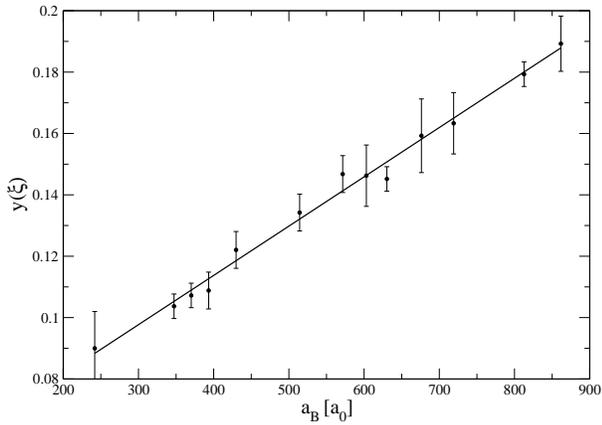}
    \caption{The universal function $y(\xi)$ as a function 
      of $a_B$ calculated from the data of Ref.~\cite{khaykovich}. The solid
line represent the best linear fit to the data.}
  \label{fig:khaykovich}
  \end{figure}
\end{center}

As a first example we analyze the binding energies of the system of three $^7$Li bosons
measured in Ref.~\cite{khaykovich} for different values
of the $^7$Li-$^7$Li scattering length. From the values of the two- and three-body
binding energies given in that reference we compute $a_B$,
the angle $\xi$ and the universal function $y(\xi)$. According to
Eq.(\ref{eq:linear})  the values of $y$ should depend   linearly
on $a_B$. The results plotted in Fig.~\ref{fig:khaykovich}
do show the expected linear behavior thus providing a further confirmation of the finite-range
theory given by Eq.(\ref{eq:energygN}). From the analysis of the straight line
we extract the values $\kappa_3^1=1.61\times 10^{-4}\;\; a_0$ and 
$\Gamma_3^1=4.95\times 10^{-2}$  (see also Ref.~\cite{gatto2013}). 

\begin{center}
  \begin{figure}[h]
    \includegraphics[width=10.0cm,angle=-90]{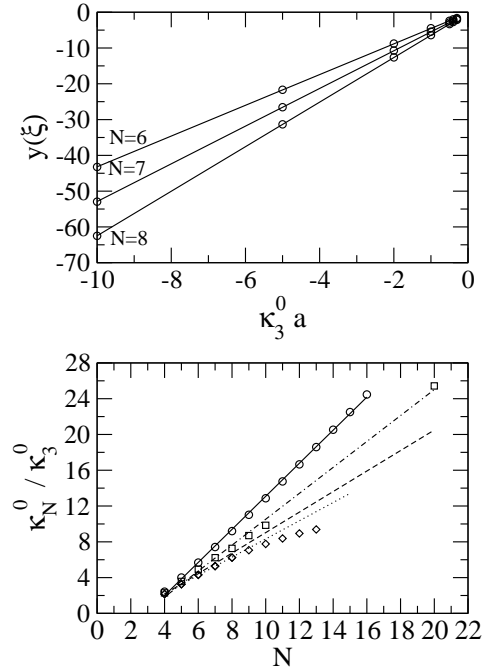}
    \caption{Analysis of the results from Ref.~\cite{stecher2010} in the
 $(a - y)$ plane (upper panel). In the lower panel the ratios
$\kappa_N^0/\kappa_3^0$
 are given  as a function of $N$ for the results of Ref.~\cite{stecher2010}
(diamonds), the present results (circles) and the results from
Ref.~\cite{blume} (squares). The universal prediction is given by the dashed
line. The solid and dashed-dotted lines are fits to the data. The dotted line is
a fit to the results of Ref.~\cite{stecher2010} up to $N=8$.}
  \label{fig:stecher}
  \end{figure}
\end{center}

As a second example we analyze the results for boson clusters obtained at the
unitary limit  by J. von Stecher in Ref.~\cite{stecher2010} using gaussian potential models
and including three-body forces. Empirically the results for $N=6,7,8$
in the  $a<0$ region were parametrized in Ref.~\cite{stecher2010} as
$E^0_N\approx(\hbar^2/m)(\kappa_N^0)^2 (x+c_Nx^{b_N})/(1+c_N)$, where
$x=(a-a^-_N)/a$. In addition, the empirical relation
$1/(\kappa_3^0 a^-_N)\approx 2.3(1)-N$ has been deduced from the numerical
result
and the ratios $(\kappa_N^0/\kappa_3^0)^2$ have been calculated up to $N=13$.
The results  from Ref.~\cite{stecher2010} cannot be plotted
in the $(a_B - y)$ plane because the explicit values of $E_2$ and
$\kappa_3^0$ were not given. Instead we can plot them in the
$(\kappa_3^0 a - y)$ plane. 
The essential characteristics of this plot, given in the upper
panel of Fig.\ref{fig:stecher}, remain the same with the
slope, given by the ratio $\kappa_N^0/\kappa_3^0$, and the shift $\Gamma_0$
related by the linear equation Eq.(\ref{eq:linear1}). 
The results lie on straight lines, as predicted by the finite-range theory
proposed here. Moreover, the straight lines seem to cross each other in one single point 
and, accordingly can be described using a global fit with $\Gamma_0\approx -0.22$.
This shift is negative and, in absolute value, it is slightly smaller
than the value obtained here for $\Gamma_0$ in the global fit.
In the lower panel of the figure the ratios $\kappa_N^0/\kappa_3^0$
are given as a function of $N$. 
The diamonds are the results from  Ref.~\cite{stecher2010}. At $N \leq 8$ they follow a linear behavior
presented by a dotted line fitted over this region. They  lie below the universal line $1+1.147(N-3)$ shown by the dashed line. On the other hand, our results, fitted by a solid line, are above the line of universality. The difference between our calculations and those  from  Ref.~\cite{stecher2010}  is due to the absence of the repulsive three-body force in our case. 
 With a two-body finite-range (gaussian) force only, the $N$-boson clusters are more bound than in the zero-range case. Including three-body repulsive force
reduces the strength  of the linear dependence between $\kappa^0_N/\kappa^0_3$ and $N$  and eventually can reproduce the universal slope of 1.147. 
The particular three-body force, selected
in Ref.~\cite{stecher2010} produces a ratio slightly lower $\kappa_4^0/\kappa_3^0$ than the universal one and, therefore, the slope of the straight dotted line in the bottom panel of
Fig. 9 is smaller, which corresponds to less bound clusters.
The calculations of Ref.~\cite{stecher2010}  deviate from the  linear behavior (dotted lines) for $N > 8$ and practically follow the square root law. This is discussed below.



The lower panel of Fig.\ref{fig:stecher} also presents the results for $N$-boson clusters of polarized tritium 
obtained in Ref.~\cite{blume} from a theoretical study within Diffusion Monte Carlo (DMC) method. In that work, the strength of the interaction has been varied  to explore a 
wide range of the $a^{-1} - K$ plane. The analysis of their results in the 
in the $a_B - y$ plane revealed that they lie on straight lines as predicted by the finite-range 
theory. Here we show the linear dependence of the energy wave numbers
on $N$ at the unitary limit. This is illustrated by the squares
whereas the dotted-dashed line represents a linear fit to the data.
 
\begin{center}
  \begin{figure}[h]
    \includegraphics[width=7.0cm,angle=-90]{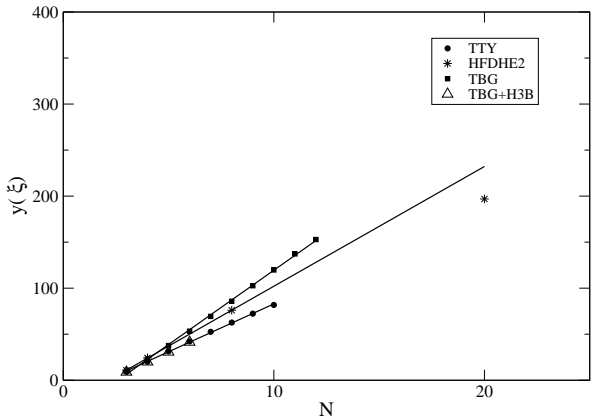}
    \caption{The universal function $y(\xi)$ as a function of $N$
calculated using the results of Ref.~\cite{lewerenz} for the TTY
potential (circles) and of Ref.~\cite{pandha1983} for the HFDHE2 potential
(stars), the present results for the TBG interaction (squares)
and the results of Ref.~\cite{gatto2012}
using a two-boy plus a three-body force (triangles). The straight lines
are the best fit to the data.}
  \label{fig:lewerenz}
  \end{figure}
\end{center}

Finally we analyze two calculations from literature performed for the ground state helium clusters with different numbers of $N$ using hard-core He-He potentials. From the published values of $E^0_N$, it is possible to determine the angle $\xi$ 
and, from it, the values of the universal function $y(\xi)$.
From the previous discussion we expect a linear relation between $N$ and $y$,
which means a quadratic dependence between $N$ and the ground state energy.
To perform this analysis we use the results from by M.
Lewerenz~\cite{lewerenz} and by Pandharipande et al.~\cite{pandha1983}.
In the former, the ground
states up to 10 atoms have been obtained within DMC  with the TTY potential~\cite{ttypotential} as the He-He interaction. In the latter,
the GFMC method has been used with the HFDHE2 interaction of
Aziz et al.~\cite{hfdhe2}. 
The $y(\xi)$ extracted from these results are shown in Fig.\ref{fig:lewerenz}
where the black circles and stars correspond to the DMC calculations with TTY and the GFMC calculations with  HFDHE2, respectively. We compare the hard-core results with 
those obtained with the soft-core TBG potentials in the present work, shown by squares.
In addition, we show by triangles the $N\leq 6$ results from Ref.~\cite{gatto2011} obtained with  the TBG potential plus a hyperradial three-body force (H3B) the of which strength is fitted to reproduce the trimer energy given by the LM2M2 potential.
 It should be noticed that the
TTY and the LM2M2 interactions produces very close results~\cite{barletta2001}
for the helium dimer and trimer and, therefore, the results of the TBG+H3B
interaction are almost on top of the results of the TTY potential. This support the
equivalence between the soft and hard-core potentials for this kind of
states, discussed for example in Ref.~\cite{kievsky2011}.

All above analyses show that the finite-range-corrected universality relations proposed 
in the present work can be used to analyze different types of
measurements and theoretical descriptions of shallow states in bosonic systems.
However, we notice that
the results obtained with HFDHE2 and plotted in Fig.\ref{fig:lewerenz}  do not follow any more
the linear behavior around $N=20$. They tend to follow a square root behavior since at
  $N\rightarrow\infty$ the ratio $E^0_N/N$ is almost constant indicating the
well known linear dependence of the energy ground state with $N$. 
It is interesting to notice that this analysis shows the transition between
the quadratic and linear behavior of the ground state energy with $N$.
In the quadratic regime the bosons are well apart, the details of the
interaction are not important and the system shows the universal behavior 
described by Eq.(\ref{eq:energygN}). As the number of bosons increases
the system becomes  more compact loosing the universal behavior. The 
strong short-range repulsion prevents the collapse of the system and
is responsible for the linear regime. This explanation clarifies
the behavior of the results from Ref.~\cite{stecher2010} for $N>8$. In this
case a very long-range repulsive three-body force has been included
making the transition to the linear regime already at low values of $N$. 
The observed transition  from the quadratic  to $E(N)$ linear behavior  for the case of realistic interaction with a (positive) large two-body
scattering lengths is very interesting and  in future should undergo a deeper analysis.


\section{Conclusions}

In the present work we have discussed DSI in shallow states of $N$-boson systems. First of all, 
we have proposed an extension of universal relation, known from three-body zero-range theory, 
to $N>3$ using the results of Refs.~\cite{hammer2007,stecher2009,deltuva2011,gatto2012}. 
This is summarized in Eq.(\ref{eq:energygN1})
which shows explicitly DSI and the two-level $N$-body structure attached to each
level $n$ of the three-boson system. Then we have extended Eq.(\ref{eq:energygN1}) to
include corrections due to the finite-range nature of the two-body interactions. 
These corrections are encoded in the shift $\Gamma_N^m$ which we introduce into Eq.(\ref{eq:energygN})
following Ref.~\cite{gatto2013}. At the origin of the extension of the universal equations to general 
$N$ is our finding that the universal function in the r.h.s. of Eq. (\ref{eq:energygN}) that 
governs the three-boson dynamics is the same as the one that governs the $N$-boson dynamics. 
Such a conclusion has been made by analyzing the solutions of the Schr\"odinger equations obtained in 
Refs.~\cite{gatto2012, gatto2013,kievsky2013,garrido2013} with various model potentials.
To our knowledge this remarkable fact was not anticipated before.

The second striking property of the universal equations (\ref{eq:energygN}) is the linear relation between
$y(\xi)$, a particular form of the universal function, and $a_B$ (or $a$ in
the zero-range theory). In the zero-range limit the straight
lines go through the origin and, accordingly, the theory has only one parameter:
the slope. In the case of finite-range interactions the linear relation
between $y(\xi)$ and $a_B$ still holds but the straight lines do not
go through the origin and, therefore, the theory has two parameters:
the slope and the distance along the $y$ axis to the origin given by
the shift $\Gamma^m_N$. This linear behavior follows from the linear
dependence of the scale parameter $\kappa^{n,m}_N$ and the finite-size
scale parameter $\Gamma^{n,m}_N$ on $N$. This property  allows us to recover the universal $N$-body 
relations in the zero-range limit.

We have checked if results available in the literature exhibit universal behavior given 
by Eq. (\ref{eq:energygN}). We have plotted in the $(a_B-y)$ plane 
the experimental data on the energies of the $^7$Li trimers, measured in 
Ref.~\cite{khaykovich} and proved that they show linear behavior. From the linear plot
we extracted the values of $\kappa_3^0$ and $\Gamma_0$. At the unitary limit,
we obtain $\kappa_3^1=1.61\times 10^{-4}\;\; a_0$,
which can be checked either experimentally or theoretically.
Furthermore, we have analyzed $N$-body energies calculated in Refs.~\cite{stecher2010,blume}
where a particular parametrization have been proposed. However, we have shown
that in both cases the linear behavior in the $(a_B-y)$ plane persists thus restricting the 
number of parameters, needed to describe these systems, to two 
for each $N$. Moreover those results show the expected linear dependence of
the energy wave number on $N$ at the unitary limit. 
All these findings support both the zero-range and finite-range universal relations
proposed here. 

A further analysis shows that universal relations persist only for attractive two-body potentials. In real systems, a  strong  repulsion between atoms an short distances exists,
which for fixed and (positive) finite values of $a_B$ leads to transition to the square root behavior of energy wave number with $N$. For particular case of Ref.~\cite{pandha1983}
this occurs around $N = 20$.
At larger $N$ the wave numbers cannot be described by the zero-range theory which
predicts a quadratic behavior of the ground state energy with $N$. In this
respect it would be  interesting to  study the boson clusters which  have negative
values of the two-body scattering length. In this case the shallow states are
better realized and the transition between the quadratic and linear regime
should occur at larger values on $N$. In fact the results of
Ref.~\cite{blume} at the unitary limit show the transition at bigger values of
$N$ than the results of Ref.~\cite{pandha1983} obtained at positive values of $a_B$.
We hope that these findings will stimulate new experimental and theoretical 
studies of $N$-body shallow states.



\end{document}